\pdfoutput=1
\documentclass[prm,superscriptaddress,amsmath,amssymb,reprint]{revtex4-2}
\usepackage{graphicx}
\usepackage{dcolumn}
\usepackage{bm}
\usepackage{wrapfig}
\usepackage{tikz}
\usepackage{graphicx}
\usepackage{subfigure}
\usepackage{blindtext}
\usepackage{textcomp}
\usepackage{makecell,multirow}
\usepackage{amsmath}
\usepackage{amsfonts}
\usepackage{ulem}
\usepackage[version=3]{mhchem}
\newcommand{\BiSe}{\ce{Bi_2Se_3}~}

\begin{document}
\preprint{AIP/123-QED}
\title{Growth of Ultrathin Bi$_2$Se$_3$ Films by Molecular Beam Epitaxy}

\author{Saadia Nasir}
\affiliation{Department of Physics and Astronomy, University of Delaware, 217 Sharp Lab, 204 The Green, Newark DE 19716 USA}
\author{Walter J. Smith}
\author{Thomas E. Beechem}
\affiliation{School of Mechanical Engineering and Birck Nanotechnology Center, Purdue University, West Lafayette, 47907, IN, USA}
\author{Stephanie Law}
\homepage{slaw@udel.edu}
\affiliation{Department of Physics and Astronomy, University of Delaware, 217 Sharp Lab, 204 The Green, Newark DE 19716 USA}
\affiliation{Department of Materials Science and Engineering, University of Delaware, 201 DuPont Hall, 127 The Green, Newark DE 19716 USA}
\date{\today}

\begin{abstract}
Bi$_2$Se$_3$ is a widely studied 3D topological insulator having potential applications in optics, electronics, and spintronics. When the thickness of these films decrease to less than approximately 6 nm, the top and bottom surface states couple, resulting in the opening of a small gap at the Dirac point. In the 2D limit, Bi$_2$Se$_3$ may exhibit quantum spin Hall states. However, growing coalesced ultra-thin Bi$_2$Se$_3$ films with a controllable thickness and typical triangular domain morphology in the few nanometer range is challenging. Here, we explore the growth of Bi$_2$Se$_3$ films having thickness down to 4 nm on sapphire substrates using molecular beam epitaxy that were then characterized with Hall measurements, atomic force microscopy, and Raman imaging. We find that substrate pre-treatment---growing and decomposing a few layers of \BiSe before the actual deposition---is critical to obtaining a completely coalesced film. In addition, higher growth rates and lower substrate temperatures led to improvement in surface roughness, in contrast to what is observed for conventional epitaxy. Overall, coalesced ultra-thin Bi$_2$Se$_3$ films with lower surface roughness enables thickness-dependent studies across the transition from a 3D-topological insulator to one with gapped surface states in the 2D regime.

\end{abstract}

\maketitle
\section{\label{sec:level1}Introduction}
Bi$_2$Se$_3$ is a widely-studied 3D topological insulator (TI) with linearly-dispersing surface states that form a single Dirac cone at the $\varGamma$ point within the bulk band gap. The surface states are spin-momentum locked, reducing back-scattering that can result in  long electron lifetimes and thus  high mobilities \cite{zhang2009topological,chen2009experimental,xia2009observation}. Typical TI thin films show promise in a variety of applications, including terahertz photodetectors and emitters\cite{zhang2010topological,giorgianni2016strong,hu2021bi2se3}, spintronic devices \cite{li2014electrical,dankert2015room,jamali2015giant,vobornik2011magnetic}, gas sensing \cite{liu2014surrounding}, and quantum computing\cite{paudel2013three}. When a TI film has a thickness less than approximately 6 nm, the wavefunction of the electrons on the top surface interacts with the wavefunction of the electrons on the bottom surface, opening a gap at the Dirac point \cite{leis2021lifting,zhang2010crossover,liu2010oscillatory,linder2009anomalous,sakamoto2010spectroscopic}. In addition, an ultra-thin TI film may act as a 2D quantum spin Hall (QSH) insulator \cite{zhang2010crossover,shan2010effective,jin2011topological}. In this case, the material only hosts two counterpropagating spin-polarized conducting channels around the perimeter of the 2D slab. This is in contrast to a 3D TI which hosts spin-polarized conducting channels on all surfaces of the material. A QSH insulator has quantized spin Hall conductance at zero field and is a promising candidate as a component in a system hosting Majorana fermions, which could lead to the realization of fault-tolerant quantum computers\cite{hart2014induced,shivamoggi2010majorana,haruyama2021quantum}. 

In order to explore the properties of ultrathin TI films, we first must be able to synthesize them at the wafer scale. Thin films of Bi$_2$Se$_3$ have been synthesized using molecular beam epitaxy (MBE) for many years\cite{liu2012characterization,bansal2011epitaxial,ginley2016topological,wang2021optimization}. These films have a hexagonal R$\bar{3}$m crystal structure with the unit cell along the [0001] direction. In this crystal structure, bismuth and selenium are arranged in successive layers with the stacking order Se$^1$-Bi-Se$^2$-Bi-Se$^1$ where the superscripts 1 and 2 represent two inequivalent positions of the selenium atoms. This five-layer repeated structure is known as a quintuple layer (QL) and is approximately 1 nm thick. The atoms within the QLs are covalently-bonded, but the consecutive QLs are connected to each other via weak van der Waals bonds. These van der Waals bonds makes it possible to grow Bi$_2$Se$_3$ thin films on a variety of substrates with relaxed constraints on lattice matching\cite{he2011epitaxial,chen2014molecular,ginley2021self,levy2018reduced}.

However, the weak interaction between the film and substrate makes it difficult to grow coalesced ultrathin films with defined thicknesses. In general, Bi$_2$Se$_3$ films nucleate as small domains. These domains grow in size as the film thickness increases, eventually fusing together to form a coalesced film\cite{schreyeck2013molecular,zhang2018diffusion,chubarov2021wafer}. However, for films with a thickness of a few nanometers (equivalent to a few QLs), these domains may or may not coalesce. This makes it difficult to study the properties of ultra-thin TI films at a wafer scale, as this requires coalesced films of a uniform thickness across the wafer. 
In this paper, we explore the growth of ultrathin (4 nm) films of Bi$_2$Se$_3$ on c-plane sapphire using molecular beam epitaxy. We show that substrate pretreatment (growth and decomposition of a few layers of \BiSe before the actual film deposition) is critical to the growth of coalesced films. In addition, we find that a low growth temperature and a high growth rate lead to better-coalesced films, contrary to conventional epitaxy\cite{joyce2000effect,zhang2018diffusion}. We explain our results using substrate wetting and adatom diffusion. Recipes to grow ultrathin Bi$_2$Se$_3$ films will allow more detailed exploration of two-dimensional TI states.

\section{\label{sec:level1}Experimental Methods}
Bi$_2$Se$_3$ thin films were grown using a Veeco GENxplor MBE system on 1$\times$1 cm$^2$ sapphire (0001) substrates. Our MBE system is equipped with a selenium valved cracker cell to improve the incorporation of selenium \cite{ginley2016growth}. The Se:Bi flux ratio is kept between 80-100 as measured by beam equivalent pressures using a flux gauge in the substrate position.   Substrates were outgassed in the load lock at 200\textcelsius\ for 12 hours before being transferred to the growth chamber whereupon they were heated to 650\textcelsius\ for 5 minutes.  The latter step ensures any impurities adsorbed on the surface have been removed. They were then cooled to the desired temperature for film deposition or surface treatment. All substrate temperatures were measured using a non-contact thermocouple located at the position of the substrate heater. To grow good-quality 4 nm Bi$_2$Se$_3$ films, we tried different growth methods, substrate temperatures, and growth rates on treated and untreated sapphire substrates as discussed in detail below. After the growth, samples were cooled to 200\textcelsius\ under a selenium flux and then removed from the growth chamber. Room temperature Hall effect measurements were taken in the van der Pauw configuration within fifteen minutes after the samples were removed from the MBE. Atomic force microscopy (AFM) scans and Raman spectroscopy measurements were performed on the films to assess film coalescence.

\section{\label{sec:level1}Results and Discussion}
\subsection{\label{sec:level2}Growths on un-treated Sapphire substrates}

\begin{figure}[h]
	\begin{minipage}[t]{\linewidth}
		\centering
		\subfigure{\includegraphics[scale=0.36]{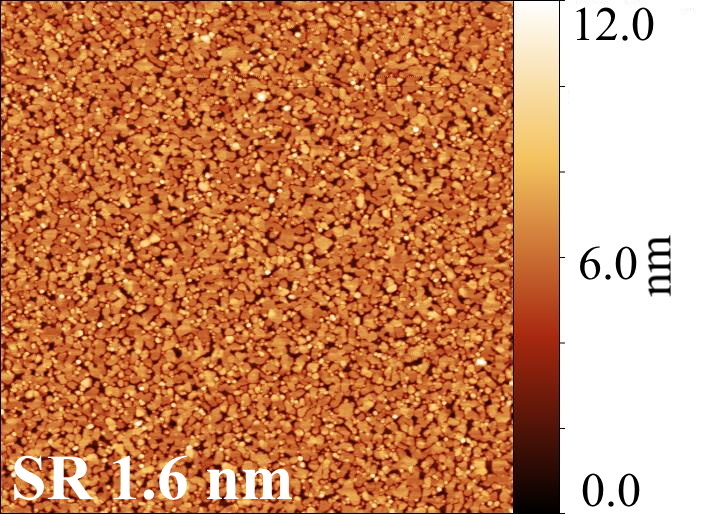}}
	\end{minipage}
	\caption{5 $\mu$m$\times$5 $\mu$m AFM scan of sample 1, a 4 nm Bi$_2$Se$_3$ thin film grown using the two-step method on an untreated sapphire substrate. The surface roughness (SR) of the sample is 1.6 nm. }
	\label{AFM1}
\end{figure}

Building from established techniques for thicker \BiSe film growth, 4 nm films were initially synthesized with a a two-step growth method at a rate of 0.60 nm/min on an untreated substrate\cite{wang2021optimization,nasir2021plane}. In the two-step method, the first few layers of the film are grown at a lower temperature and then the rest of the film is grown at a comparatively higher temperature. We grew sample 1 by co-depositing bismuth and selenium for one minute and annealing for eighty seconds with continual selenium flux at 325\textcelsius. This grow-anneal process was repeated until the desired seed layer thickness of 3 nm was reached. Then the substrate temperature was raised to the second set point (100\textcelsius\ higher than the first growth temperature unless mentioned otherwise) under a continual selenium flux to grow the remaining 1 nm of the film by simply co-depositing bismuth and selenium. Growth details for this sample---and all samples in this study---are summarized in Table \ref{table}. Hall measurements on this sample indicated that the film was insulating. FIG. \ref{AFM1} displays a $5\times5$ $\mu$m AFM image of this sample where discrete small islands are observed and the typical triangular morphology for \BiSe synthesis is not observed. The comparatively large value of surface roughness at 1.6 nm suggests a non-coalesced film.  The insulating nature of the films as determined by Hall measurements further supports this conclusion.

\begin{figure}[h]
	\begin{minipage}[t]{\linewidth}
		\centering
		\subfigure{\includegraphics[scale=0.36]{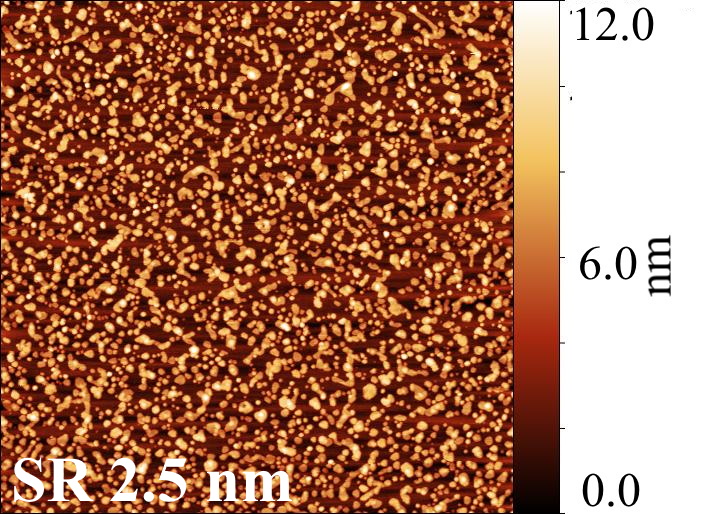}}
	\end{minipage}
	\caption{5 $\mu$m$\times$5 $\mu$m AFM scan of ample 2, a Bi$_2$Se$_3$ thin film grown directly on sapphire using the two-step method, followed by annealing for one hour under a Se flux at 425\textcelsius. The surface roughness (SR) of the sample is 2.5 nm. }
	\label{AFM2}
\end{figure}

In an attempt to realize fully coalesced films, sample 2 was grown using the same procedure as sample 1 but with an additional annealing step in a selenium atmosphere at 425\textcelsius\ for one hour after the growth was concluded. The goal of this process was to provide enough time for the bismuth and selenium atoms to diffuse across the substrate and result in a coalesced film. Contrary to our expectation, the extra annealing did not result in coalescence but instead facilitated growth of regions which are even more separated, as is evident in the  AFM image of sample 2 shown in FIG. \ref{AFM2}. Quantitatively, the surface roughness of the sample increased to 2.5 nm and the Hall measurement showed insulating behavior. Taken in aggregate, we therefore conclude that sample 2---even with the additional annealing step---did not coalesce.

\begin{table*}
	\caption{Growth conditions and room temperature transport properties of 4 nm Bi$_2$Se$_3$ films grown on sapphire substrates. The temperature for the second step in the two-step growth is 100\textcelsius\ higher than first step except for sample I where the temperatures are mentioned explicitly. The sheet density and mobility values listed below are the averages resulting from three consecutive Hall measurements.}
	\begin{ruledtabular}
		\begin{tabular}{ccccccc}
			
			 &Substrate &Growth&Growth&Growth& Sheet   & Mobility \\ 
			Sample&Pre-Treatment&method &temperature&rate&Density ($n_s$)&$\mu$\\
			
		&&&(\textcelsius)&(nm/min)&$x10^{13}(cm^{-2})$&$(cm^{2}/Vs)$\\
			\hline
              1&No&Two-step&325&0.6&Insulating&Insulating\\  
              2$^{*}$&No&growth&325&0.6&Insulating&Insulating\\    
              \hline
			 A&Yes&&325&0.25& -2.29 $\pm$ 0.04& -118.8 $\pm$ 3.4 \\
			 B&Yes&Direct&325&0.60& -2.91 $\pm$ 0.03& -196.0 $\pm$ 3.1 \\
			 C&Yes&growth&300&0.25& -2.90 $\pm$ 0.01& -138.1 $\pm$ 2.5 \\
		     D&Yes&&300&0.60& -2.91 $\pm$ 0.005& -177.8 $\pm$ 2.5 \\
		     \hline
			 E$^{**}$&Yes&&325&0.25& -2.30 $\pm$ 0.26& -69.4 $\pm$ 6.7 \\
			 F&Yes&Two-&325&0.60& -2.81 $\pm$ 0.005& -141.8 $\pm$ 1.8 \\
		     G&Yes&step&300&0.25& -2.53 $\pm$ 0.05& -97.8 $\pm$ 2.3 \\
		      H$^{***}$&Yes&growth&300&0.60&-&-\\
		      I&Yes&&325(3 nm)/400(1 nm)&0.60&-2.61 $\pm$ 0.11&-168.9 $\pm$ 6.2
		\end{tabular}
		\label{table}
	\end{ruledtabular}\\
	* Sample 2 was annealed in selenium at 425\textcelsius\ for 1 hr after the two-step growth.\\
        ** For sample E, the Hall voltage vs. magnetic field curve was nonlinear and inconsistent.\\
	*** The film appeared non-uniform close to one corner; we could not make good contact to obtain Hall data. 
\end{table*}

\begin{figure*}[!ht]
	\begin{minipage}[h]{\linewidth}
		\centering
		\subfigure{\includegraphics[trim={0cm 0cm 6cm 0cm},clip,scale=0.64]{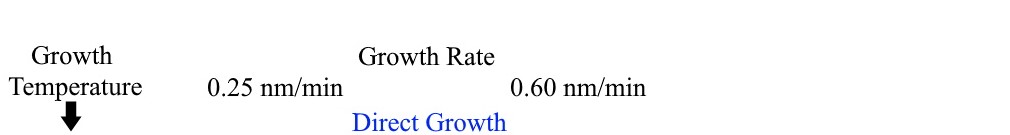}}
		\subfigure{\includegraphics[trim={0cm 0cm 2cm 0cm},clip,scale=0.64]{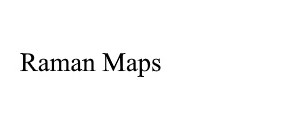}}\\
		\subfigure{\includegraphics[trim={1cm 0cm 0.25cm 0cm},clip,scale=0.65]{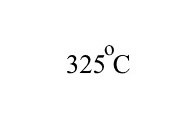}}
		\subfigure{\includegraphics[scale=0.265]{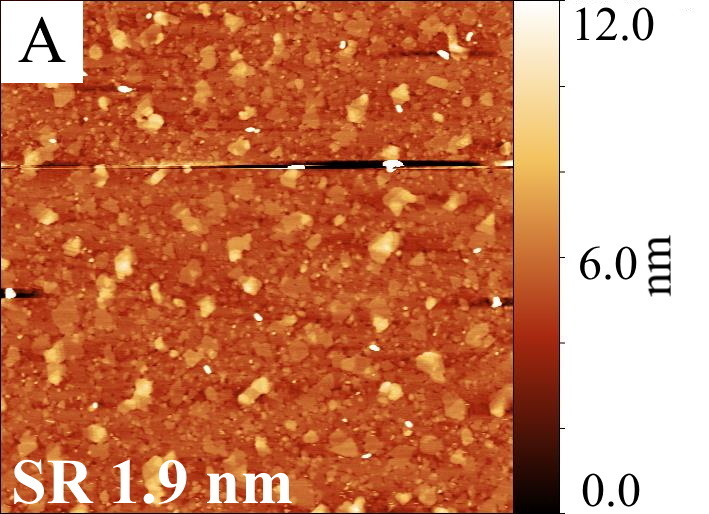}}\,\,
		\subfigure{\includegraphics[scale=0.265]{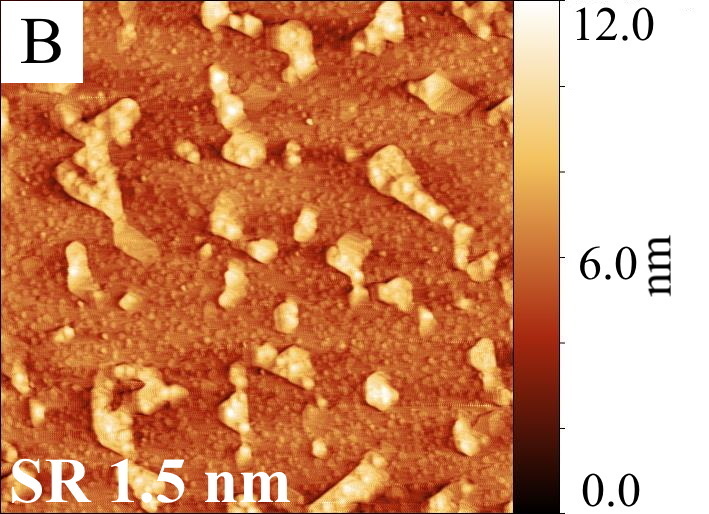}}
		\subfigure{\includegraphics[scale=0.256]{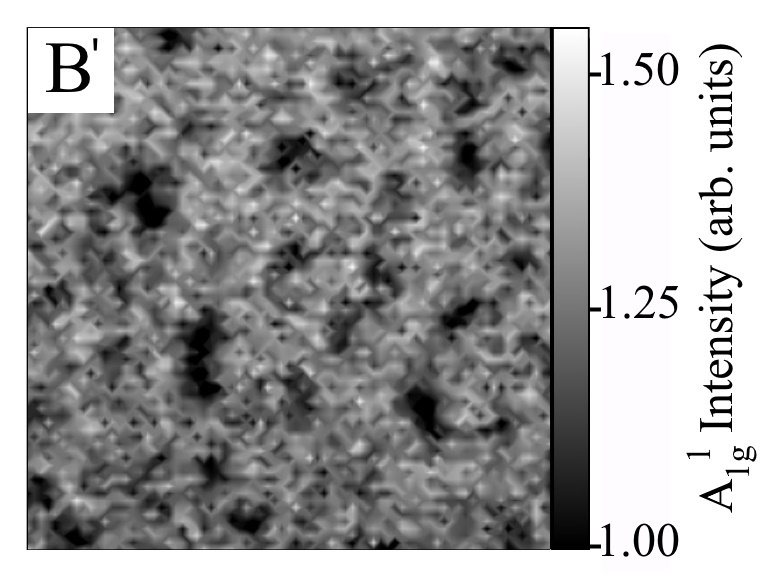}}\\
		\subfigure{\includegraphics[trim={1cm 0cm 0.5cm 0cm},clip,scale=0.7]{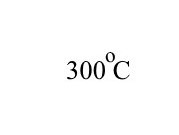}}
    	\subfigure{\includegraphics[scale=0.265]{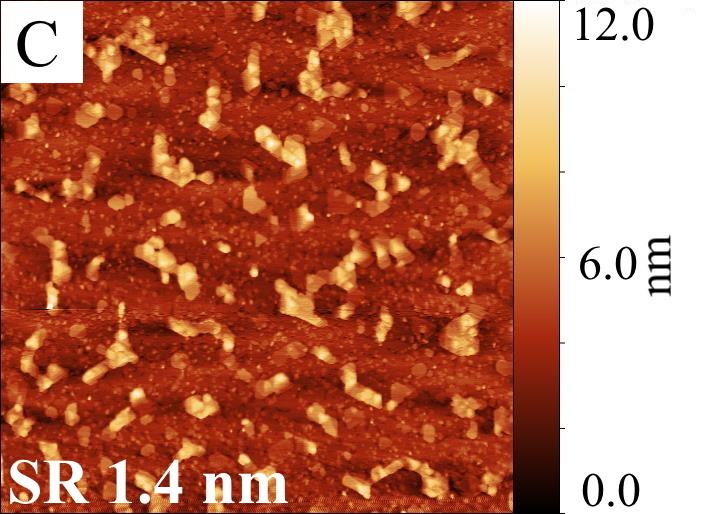}}\,\,
        \subfigure{\includegraphics[scale=0.265]{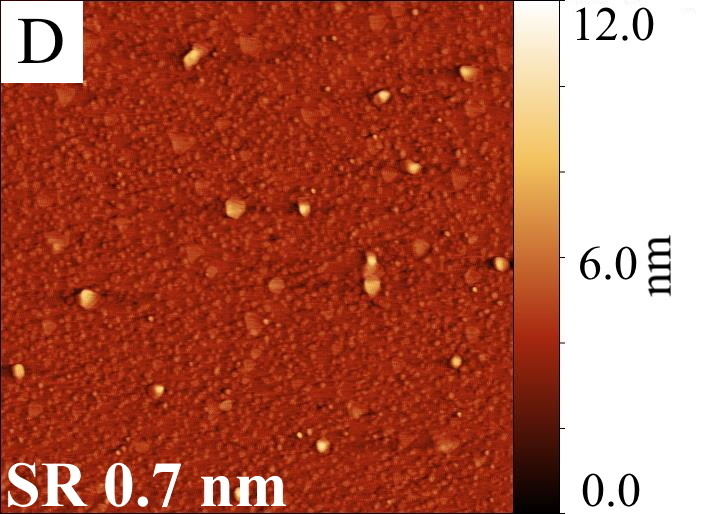}}
		\subfigure{\includegraphics[scale=0.256]{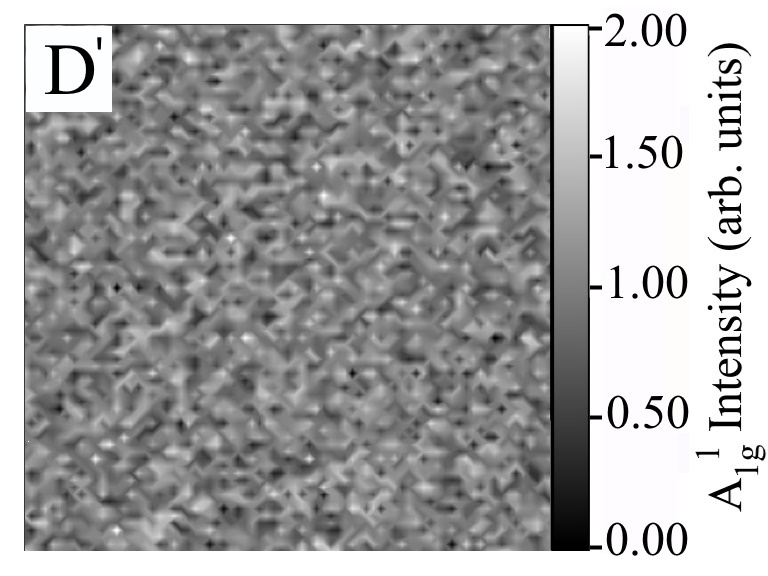}}\\
		\subfigure{\includegraphics[trim={1cm 1cm 0cm 1.2cm},clip,scale=0.64]{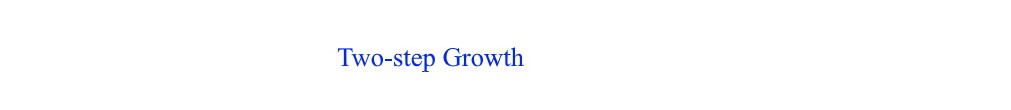}}
		\subfigure{\includegraphics[trim={1cm 0cm 0.25cm 0cm},clip,scale=0.65]{GT2.jpg}}
		\subfigure{\includegraphics[scale=0.265]{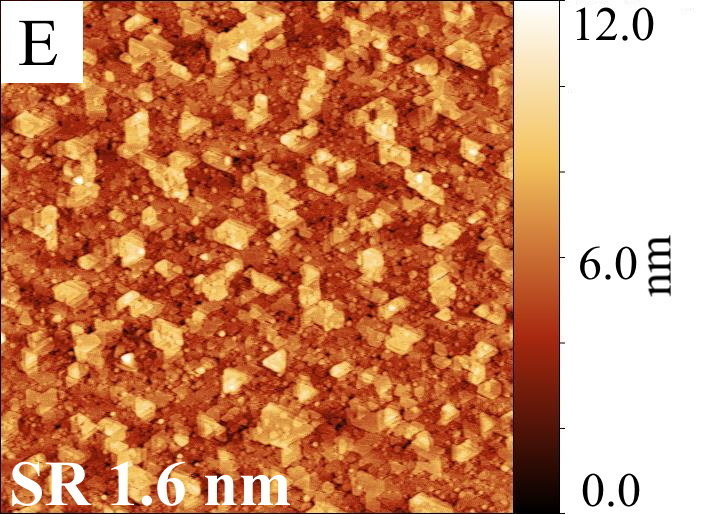}}\,\,
		\subfigure{\includegraphics[scale=0.265]{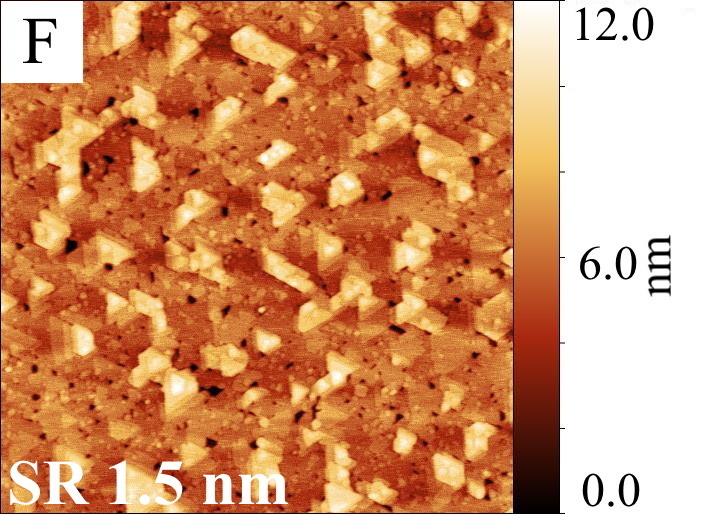}}
		\subfigure{\includegraphics[scale=0.256]{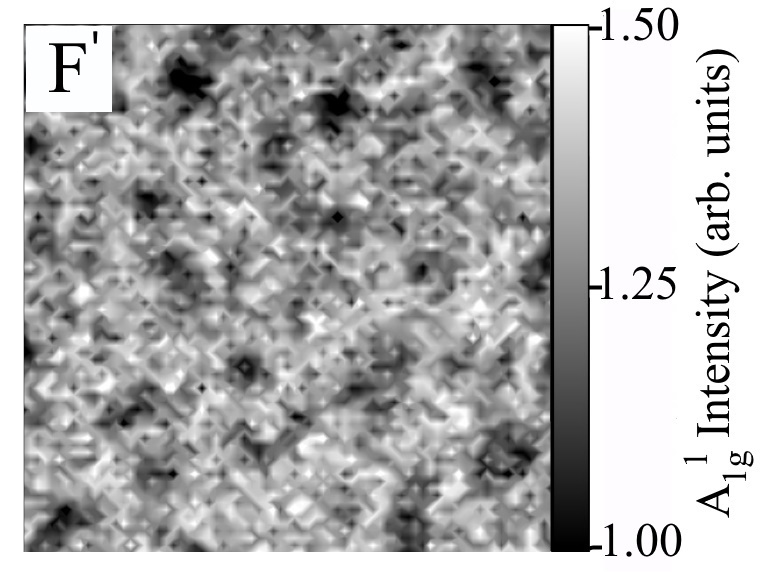}}\\
		\subfigure{\includegraphics[trim={1cm 0cm 0.5cm 0cm},clip,scale=0.7]{GT3.jpg}}	
		\subfigure{\includegraphics[scale=0.265]{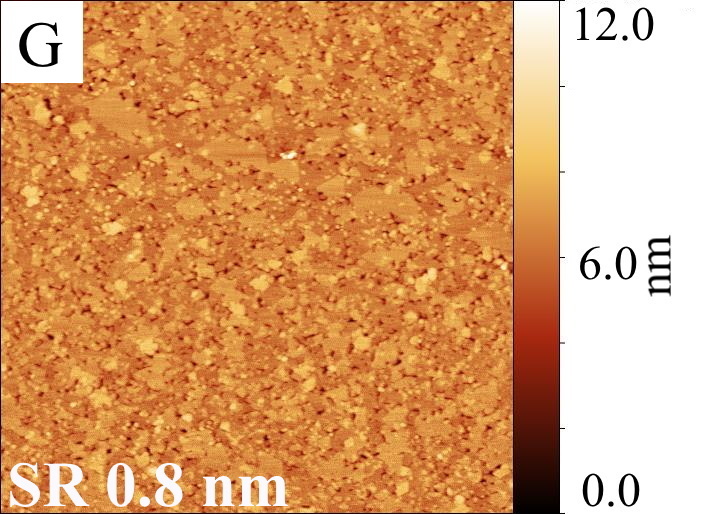}}\,\,
		\subfigure{\includegraphics[scale=0.265]{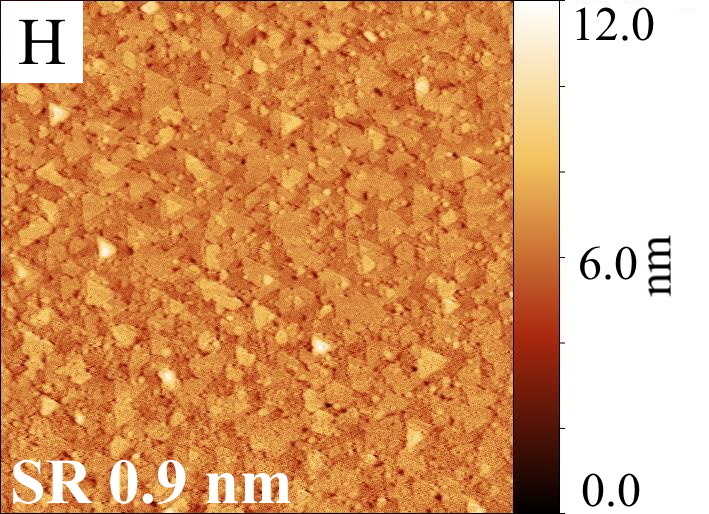}}
		\subfigure{\includegraphics[scale=0.256]{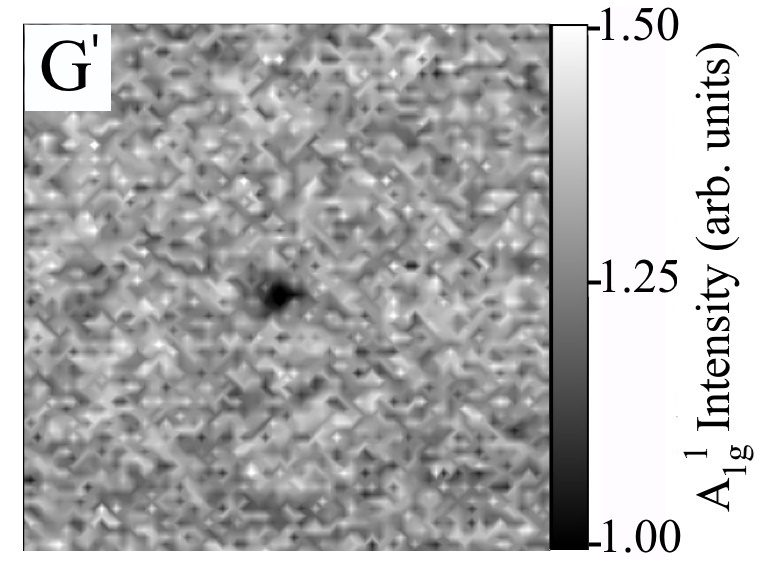}}
	\end{minipage}
	\caption{5 $\mu$m$\times$5 $\mu$m AFM scans (red color scale; left and middle columns) and Raman maps (grayscale, right column) of the 4 nm Bi$_2$Se$_3$ films. The sample identifiers are given in the top left corner and are the same as in Table \ref{table}. A prime is added to the sample names in Raman maps. The surface roughness of samples is given on the bottom left of the AFM images. All samples were grown on pretreated sapphire substrates. Samples A-D are grown using the direct growth method and samples E-H are deposited by the two-step growth method. Growth rates are given above the respective AFM columns and growth temperatures are given at the left of each row. Raman and AFM images do not correspond to the same location on a given sample.}
\label{AFM3}	
\end{figure*}
\subsection{\label{sec:level2}Growths on pre-treated Sapphire substrates}
Based on the results of the first two growths, we hypothesized that Bi$_2$Se$_3$ was not wetting the sapphire substrate, causing island-type growth and precluding film coalescence upon annealing. To overcome this issue, we decided to change the chemistry of the substrate surface through pre-treatment. Pre-treatment of sapphire substrates has been used to improve the growth of other van der Waals materials including \BiSe,Bi$_2$Te$_3$, Ga$_2$Se$_2$ and WSe$_2$ \cite{levy2018reduced,chegwidden1998molecular,mortelmans2019peculiar}. Although the mechanism is not entirely clear, it is suspected that the pre-treatment may passivate dangling bonds and step edges, making it easier for the film to wet the substrate \cite{kampmeier2015suppressing}. 

For all subsequent samples, a 5 nm Bi$_2$Se$_3$ thin film was grown on the sapphire substrate by first co-depositing bismuth and selenium at a single growth temperature of 325\textcelsius\ with no annealing. The samples were then heated in a selenium flux to 470\textcelsius\ , which is higher than the Bi$_2$Se$_3$ thermal decomposition temperature for our MBE. We kept the samples at 470\textcelsius\ for 30 minutes to completely desorb the Bi$_2$Se$_3$ film, as was confirmed by monitoring the reflection high energy electron diffraction (RHEED) pattern. Specifically, at the end of this step, the RHEED pattern only showed lines associated with the sapphire substrate. After Bi$_2$Se$_3$ decomposition, the substrates were cooled to the required growth temperatures. Two sets of films were then synthesized on the pre-treated substrates.  First, Samples A-D were grown using direct co-deposition of bismuth and selenium at a given temperature with no annealing. Substrate temperatures of 300\textcelsius\ or 325\textcelsius\ and growth rates of 0.25 or 0.6 nm/min were employed for this direct growth process. Second, Samples E-I were grown using the two-step growth technique analogous to Sample 1. Initial growth temperatures of 300\textcelsius\ and 325\textcelsius\ were tried and growth rates of both 0.25 nm/min and 0.6 nm/min were employed. The specific growth rates and growth temperatures for Samples A-I along with their transport properties are shown in Table \ref{table}.

All samples grown on the pre-treated sapphire substrate were electrically conductive, suggesting a degree of coalescence for these films that at least reached the percolation threshold. AFM and Raman images provided in Fig. \ref{AFM3} support this conclusion when analyzed in tandem. False color Raman images were created by assigning color scaled to the strength of the $A_{1g}^2$ mode of Bi$_2$Se$_3$ (for details, see Supplemental Material \cite{supp}).  Shown in the rightmost column of FIG. \ref{AFM3} and indicated by the sample letter with a prime, brighter regions correspond to stronger \BiSe signals while darker regions are indicative of the inverse. Importantly,  modes stemming from \BiSe are present in all spectra\textemdash both bright and dark regions\textemdash consistent with a coalesced film. By comparing the AFM images and Raman maps in FIG. \ref{AFM3}, the number and distribution of dark spots found in the Raman images are seen to roughly correlate with the distribution of columnar regions observed in the AFM images (\textit{i.e.,~} ``bright regions" in topographic images).  Simply put, columnar regions seem correlated with regions of lower Raman signal.  Taken in aggregate, we therefore conclude that \BiSe is present over the entirety of the scanned range and not just within columnar ``islands."

Despite the common film coalescence, samples grown on pre-treated substrates exhibit significantly different morphologies. Changes in morphology, in turn, provide insight into the growth mechanisms at play. We can first compare pairs of films with different growth rates by looking across the rows in FIG. \ref{AFM3}: sample A vs. B, C vs. D, E vs. F, and G vs. H. We generally observe a lower surface roughness and larger domains for the samples grown with a higher growth rate, contrary to expectations. Next, we can compare films grown at different temperatures by looking down the columns in FIG. \ref{AFM3}: sample A vs. C, B vs. D, E vs. G, and F vs. H. We see a lower surface roughness for the samples grown at a lower substrate temperature, again contrary to expectations. To determine the optimal substrate temperature, we grew an additional sample using the two-step growth method in which the first set point was reduced to 275\textcelsius. However, this did not noticeably improve the surface roughness and resulted in smaller domains. The AFM image for this sample is given in the Supplemental Material FIG. S2 \cite{supp}. Finally, we can compare samples grown with the direct growth method and those grown with the two-step growth method: sample A vs. E, B vs. F, C vs. G, and D vs. H. Although the surface roughness is similar for both methods, the morphology of the films grown with the two-step method is more typical of what is observed for thicker \BiSe films. For the films grown with the two-step method, we see the typical triangular domain morphology which we interpret to be indicative of better-quality films. 

\begin{figure}[h]
	\begin{minipage}[t]{\linewidth}
		\centering
		\subfigure{\includegraphics[scale=0.36]{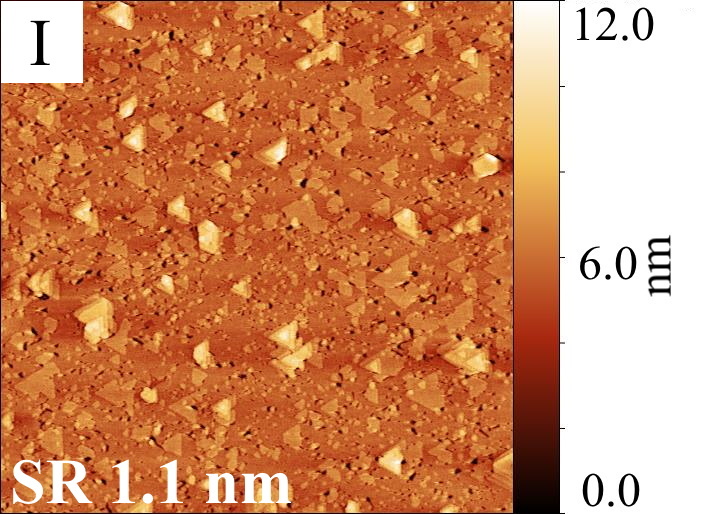}}
	\end{minipage}
	\caption{5 $\mu$m$\times$5 $\mu$m AFM scan of a Bi$_2$Se$_3$ thin film grown on treated sapphire by two-step method with growth rate of 0.6 nm/min, using growth temperatures of 325\textcelsius\ for the first step and 400\textcelsius\ for the second step. The surface roughness (SR) is 1.1 nm.}
	\label{AFM4}
\end{figure}

To investigate if we could further optimize the growth by adjusting the temperatures of the first and second steps, we grew a final sample on a treated substrate using the two-step growth method with a growth rate of 0.6 nm/min. The first 3 nm of the film were grown at a substrate temperature of 325\textcelsius\ while the final 1 nm was grown 75\textcelsius\ higher at 400\textcelsius. The AFM image of the resulting sample I is shown in FIG. \ref{AFM4}. We can compare this sample with both sample F (same initial step temperature) and sample H (same final step temperature). Sample I has a lower RMS roughness than sample F but a higher RMS roughness than sample H, indicating that the final step temperature has more influence on the surface roughness than the initial step temperature.

\begin{figure}[h]
	\begin{minipage}[t]{\linewidth}
		\centering
		\subfigure{\includegraphics[trim={0.8cm 0cm 0cm 0cm},clip,scale=0.35]{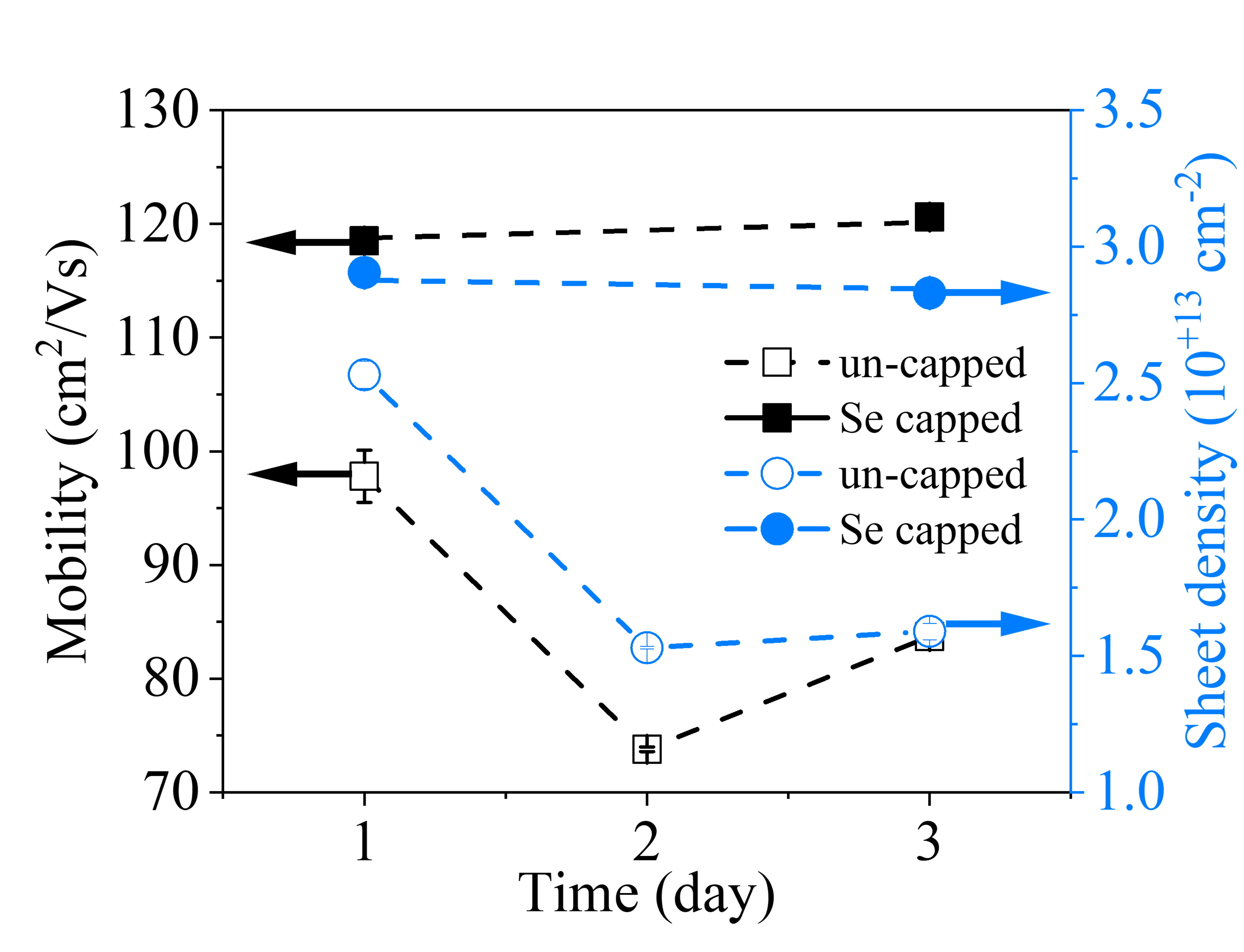}}
	\end{minipage}
	\caption{Mobility and sheet density of Se capped and un-capped 4 nm Bi$_2$Se$_3$ film grown on a treated sapphire substrate, using two-step growth method with growth temperature of 300\textcelsius\ and growth rate of 0.25 nm/min. Error bars for the mobility and sheet densities are smaller than the symbol sizes hence can not be seen.}
	\label{hall}
\end{figure}


As described above, we used room-temperature Hall effect measurements to understand the transport properties of the films. The sheet densities are similar for all samples A-I regardless of the growth methods and conditions with only small changes as the film morphology changes. We attribute this consistency to defects at the film/substrate interface which dominate the transport properties. These defects can be reduced by growing a buffer layer between the sapphire substrate and the Bi$_2$Se$_3$ thin films\cite{wang2018growth,wang2011superlattices,chen2014molecular}. Unlike the sheet density, the mobility changes with growth method. The mobility in these ultrathin films will always be inherently limited by scattering from the top and bottom surfaces, but it can also be limited by surface roughness and scattering from grain boundaries. We find that the mobility increases with increasing growth rate, which is consistent with the reduction in roughness observed in the AFM images. The effect of growth temperature on the mobility is smaller and not consistent across all pairs of samples. Samples B and C show a relatively high mobility despite their columnar structure. In these samples, the electrons may be traveling through the thin coalesced "background" film rather than through the tall islands. 
\subsection{\label{sec:level2}Selenium capping on ultrathin \BiSe film}
Additionally, we note  that the uncapped films like those described in Table \ref{table} and FIG. \ref{AFM3} showed severe aging effects. To reduce this effect, an additional sample was grown using conditions identical to Sample G and capped with selenium\cite{lin2018new}. To do this, before taking the sample out of growth chamber, we cooled it from 200\textcelsius\ to 65\textcelsius\ in vacuum and then exposed it to the Se flux for forty minutes, resulting in a crystalline Se capping layer as shown in the Supplemental Material FIG. S3 \cite{supp}. It can be seen in FIG. \ref{hall} that the selenium capping layer resulted in much more reproducible transport measurements over time compared to the uncapped sample. The increase in sheet density for the capped sample is likely caused by a change in band bending due to the capping layer, while the increase in mobility is likely due to a decrease in oxidation and surface adsorbates.


\section{\label{sec:level1}Summary and Conclusion}
Overall, we found that for growth on pre-treated substrates, a faster growth rate and a lower growth temperature produced smoother films. We also found that the two-step growth method resulted in the usual triangular domain film morphology. These results are consistent with the aforementioned hypothesis that Bi$_2$Se$_3$ does not wet the sapphire substrate well. Deposition at a high growth rate and a low substrate temperature reduces adatom mobility. In growth of normal covalently-bonded materials, this would lead to rougher surfaces. However, for Bi$_2$Se$_3$ grown by van der Waals epitaxy on sapphire, the low adatom mobility prevents the film from forming islands to minimize its contact with the substrate. When adatom mobility is high, the bismuth atoms are able to diffuse along the substrate, find an existing Bi$_2$Se$_3$ domain, diffuse up the domain sidewalls, and incorporate on top, leading to islands or columns with reduced substrate coverage. By reducing the adatom mobility, we can limit this behavior and induce the film to nucleate across the entire substrate. The two-step growth method further improves the film morphology by increasing adatom mobility once the film has nucleated on the substrate. This is, of course, all predicated on using the substrate pre-treatment to improve the wettability of \BiSe on the substrate. 

We remain unsure of precisely how the substrate pre-treatment is changing the surface chemistry. However, we can make an educated guess by looking at similar systems \cite{chegwidden1998molecular,kampmeier2015suppressing}. In previous experiments, x-ray photoemission spectroscopy (XPS) was used to investigate sapphire substrates that were pre-treated with GaSe in a similar manner to our procedure. They found that after the Ga-Se film was desorbed from the substrate, both gallium and selenium peaks were still visible in XPS measurements. They hypothesized the existence of a reacted layer in which gallium and selenium were bonded to the sapphire substrate. In this reacted layer, selenium may have replaced oxygen in the substrate, perhaps making the surface less polar, less reactive, and therefore more easily wetted. It is also possible that the adatoms are incorporating at step edges and passivating surface dangling bonds without atomic replacement. It is likely that similar reactions are happening in our samples, but further work is needed to fully understand the surface chemistry of the substrate after pre-treatment.  

In summary, we grew ultra-thin Bi$_2$Se$_3$ films using a variety of different growth recipes and conditions on treated and un-treated sapphire (0001) substrates. We demonstrated that pre-treatment of the sapphire substrate results in better substrate coverage and improvement in domain coalescence. We also demonstrated that the two-step growth method on the pre-treated substrate results in the typical triangular domain morphology of Bi$_2$Se$_3$. We observed an improvement of the surface roughness and the film morphology with higher growth rates and lower substrate temperatures, contrary to that usually observed for epitaxial growths of conventional materials. By capping the thin films with crystalline selenium, their surface quality can be preserved. It makes these materials reliable for the study of thickness-dependent optical or electronic properties around the critical thickness of 6 nm and can provide us with useful information for device applications.


\begin{acknowledgments}
S. N. and S. L acknowledge funding from the U.S. Department of Energy, Office of Science, Office of Basic Energy Sciences, under Award No. DE-SC0017801. The authors acknowledge the use of the Materials Growth Facility (MGF) at the University of Delaware, which is partially supported by the National Science Foundation Major Research Instrumentation under Grant No. 1828141 and UD-CHARM, a National Science Foundation MRSEC under Award No. DMR-2011824.
\end{acknowledgments}

\section*{Data Availability}
The data used in this paper can be provided upon reasonable request.
\bibliography{bibliography}
\end{document}


\preprint{AIP/123-QED}
\title{Supplemental Material: Growth of Ultrathin Bi$_2$Se$_3$ Films by Molecular Beam Epitaxy}

\author{Saadia Nasir}
\affiliation{Department of Physics and Astronomy, University of Delaware, 217 Sharp Lab, 204 The Green, Newark DE 19716 USA}
\author{Walter J. Smith}
\author{Thomas E. Beechem}
\affiliation{School of Mechanical Engineering and Birck Nanotechnology Center, Purdue University, West Lafayette, 47907, IN, USA}
\author{Stephanie Law}
\homepage{slaw@udel.edu}
\affiliation{Department of Physics and Astronomy, University of Delaware, 217 Sharp Lab, 204 The Green, Newark DE 19716 USA}
\affiliation{Department of Materials Science and Engineering, University of Delaware, 201 DuPont Hall, 127 The Green, Newark DE 19716 USA}
\date{\today}
\maketitle
\onecolumngrid
 
\begin{figure*}[h]
	\begin{minipage}[t]{\linewidth}
		\centering
		\subfigure{\includegraphics[scale=0.8]{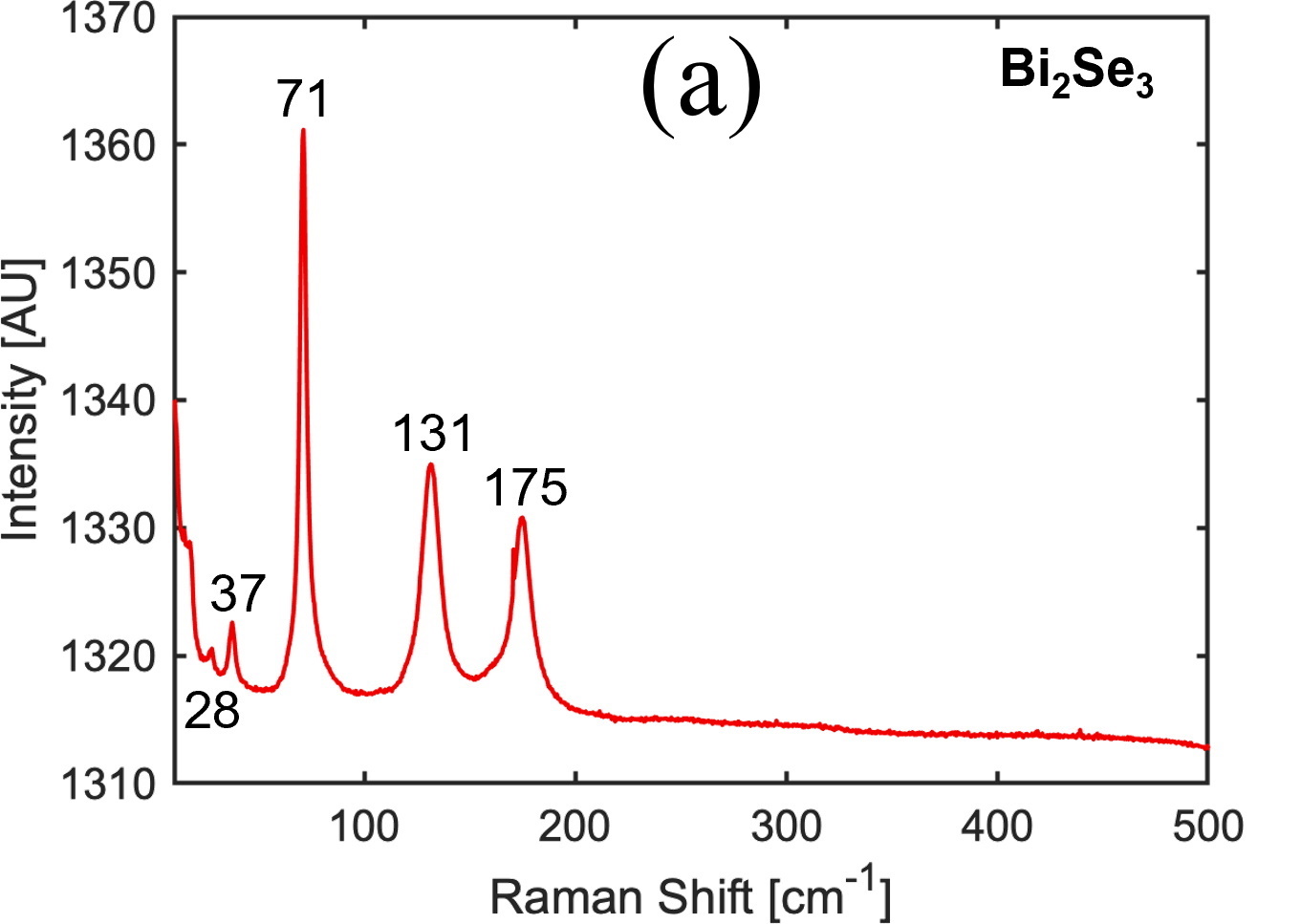}}
		\subfigure{\includegraphics[scale=0.8]{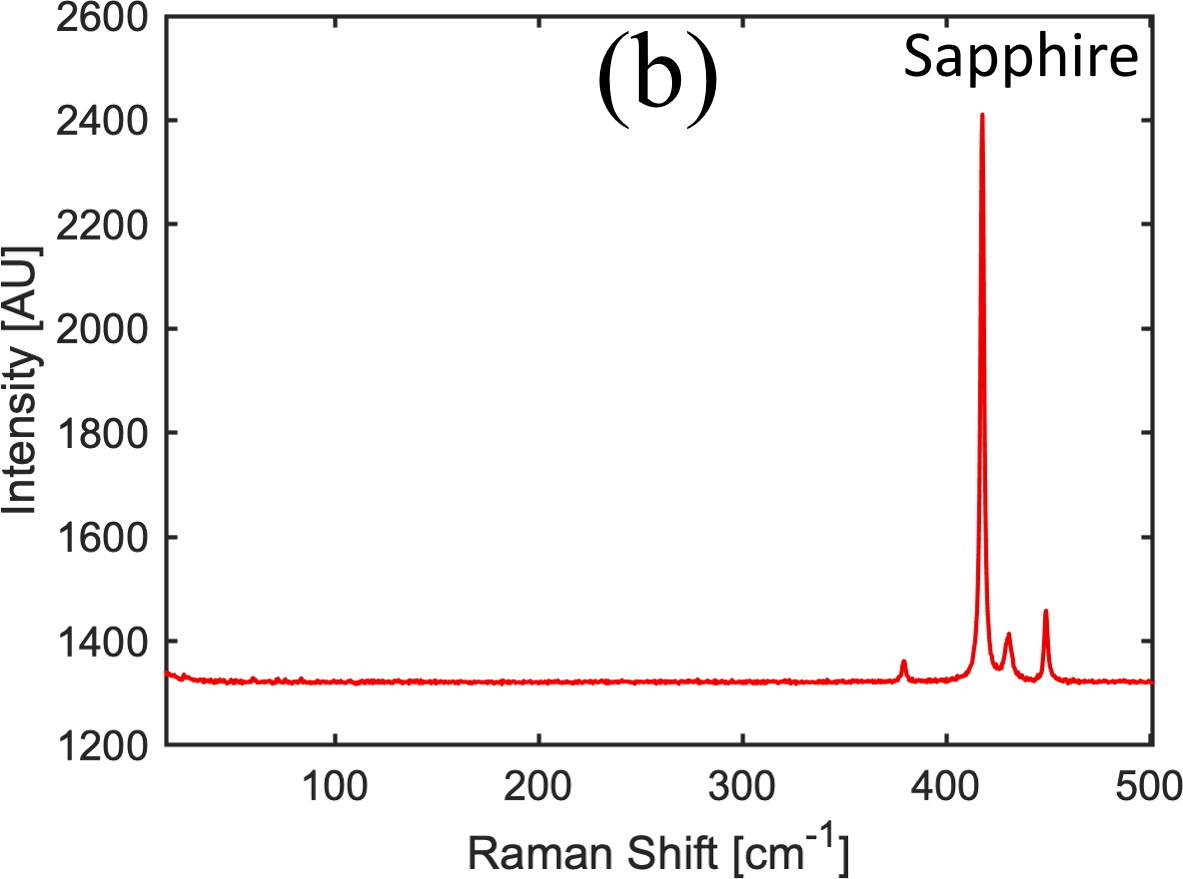}}
	\end{minipage}
	\caption{Raman spectra of a) Bi$_2$Se$_3$ thin film and b) sapphire.}
	\label{Raman}
\end{figure*}

All Raman measurements were performed using 0.5 mW of 532 nm laser light focused to an approximately diffraction limited spot-size of 360 nm with a 100X/0.9 NA objective in a backscattering arrangement.  Scattered light was dispersed with a Czerny-Turner spectrometer using  an 1800 l/mm grating resulting in a spectral accuracy of $<$ 1 cm-1.  Raman images were acquired by collecting spectra over 5x5 µm area with collections taking place every 67 nm.  False color Raman images were created by summing the Raman intensity of the $A_{1g}^2$ mode near 70 ${cm^{-1}}$ as shown in FIG. \ref{Raman}.   Beyond the $A_{1g}^2$, three additional primary Raman modes were observed consistent with expectation\cite{zhang2011raman}.  FIG. \ref{Raman}(b) also provides the Raman response of sapphire taken from a region of the sample masked off to prevent \BiSe deposition.  The major Raman modes of sapphire lie above above 350 cm$^{-1}$\cite{hushur2009raman}  and thus are spectrally well separated from that \BiSe allowing for easy differentiation between the materials.  The strength of the sapphire signal is, however, much smaller than that of \BiSe. To gain comparative signal strength from the sapphire, laser power was increased by 40x.  This smaller Raman cross section for sapphire can also be the reason why its response is not observed even in the ``dark spots" of Figure 3 within the main manuscript. 


 \begin{figure}[h]
	\begin{minipage}[t]{\linewidth}
		\centering
		\subfigure{\includegraphics[scale=0.5]{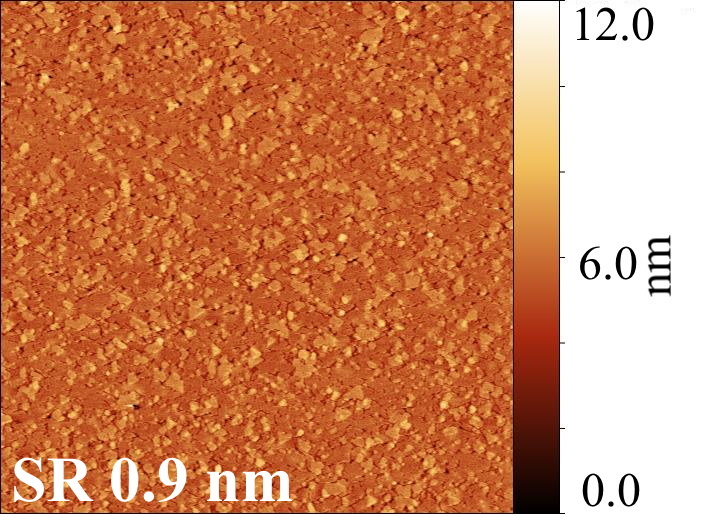}}
	\end{minipage}
	\caption{5$\times$5 $\mu$m AFM scan of a Bi$_2$Se$_3$ thin film grown on a pre-treated sapphire substrate using the two-step method with a growth rate of 0.6 nm/min and a growth temperature of 275\textcelsius\ for the first step and 375\textcelsius\ for the second step. The surface roughness (SR) is given in the bottom left corner.}
	\label{AFM2}
\end{figure}

\begin{figure}[h]
	\begin{minipage}[t]{\linewidth}
		\centering
		\subfigure{\includegraphics[scale=0.5]{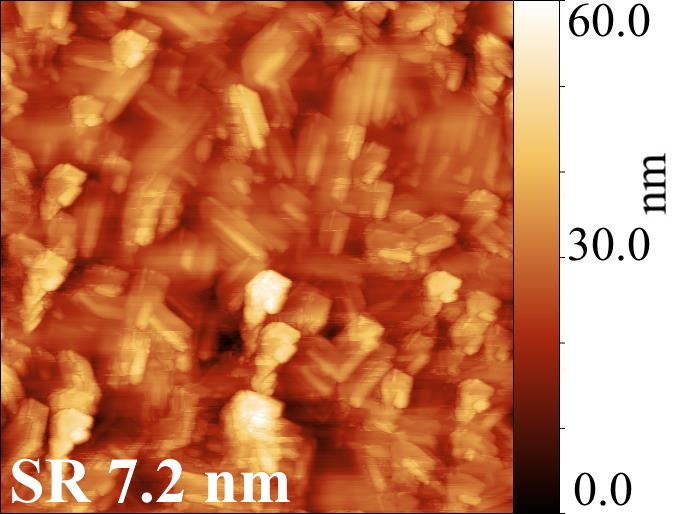}}
	\end{minipage}
	\caption{5$\times$5 $\mu$m AFM scan of a selenium-capped 4nm Bi$_2$Se$_3$ film. Thickness of the capping layer is 30-35 nm as determined using profilometry. The surface roughness (SR) is given in the bottom left corner.}
	\label{AFM4}
\end{figure}

FIG. \ref{AFM2} shows the AFM scan for the Bi$_2$Se$_3$ film grown on a pre-treated sapphire substrate using the two-step method with a growth rate of 0.6 nm/min and a growth temperature of 275\textcelsius\ for the first step and 375\textcelsius\ for the second step. We see that reducing the growth temperature further than sample H reduces domain size and does not improve the surface roughness. FIG. \ref{AFM4} shows the AFM scan for the Se-capped Bi$_2$Se$_3$ sample. This image is consistent with a polycrystalline capping layer  \cite{lin2018new}.

\clearpage
\textbf{REFERENCES}
\bibliography{bibliography}